\documentclass[10pt,journal,compsoc]{IEEEtran}

\usepackage{amssymb}
\usepackage{pdfpages}
\usepackage{enumitem}
\usepackage{bbm}
\usepackage{multirow}
\usepackage{algorithm,algpseudocode}
\usepackage{url}

\usepackage{setspace,cite}
\usepackage{multirow,multicol}
\usepackage[tbtags]{amsmath}
\usepackage{amsbsy}
\usepackage{amssymb}
\usepackage{amsfonts}

{\begin{list}               
    {$\bullet$ \hfill}{
        \setlength{\leftmargin}{\parindent}
        \setlength{\parsep}{0.04\baselineskip}
        \setlength{\itemsep}{0.5\parsep}
        \setlength{\labelwidth}{\leftmargin}
        \setlength{\labelsep}{0em}}
    }
{\end{list}}

\providecommand{\eref}[1]{\eqref{#1}}  
\providecommand{\cref}[1]{Chapter~\ref{#1}}

\providecommand{\fref}[1]{Figure~\ref{#1}}

\providecommand{\R}{\ensuremath{\mathbb{R}}}

\renewcommand{\vec}[1]{\ensuremath{\boldsymbol{#1}}}
\providecommand{\mat}[1]{\ensuremath{\boldsymbol{#1}}}

\providecommand{\calL}{\mathcal{L}}

\providecommand{\vx}{\vec{x}}

\providecommand{\mTheta}{\mat{\Theta}}

\providecommand{\veta}{\vec{\eta}}

\newcommand{\argmin}[1]{\mathop{\underset{#1}{\mbox{argmin}}}}

\usepackage{array}
\newcommand{\PreserveBackslash}[1]{\let\temp=\\#1\let\\=\temp}
\newcolumntype{C}[1]{>{\PreserveBackslash\centering}p{#1}}
\newcolumntype{R}[1]{>{\PreserveBackslash\raggedleft}p{#1}}
\newcolumntype{L}[1]{>{\PreserveBackslash\raggedright}p{#1}}

\begin{document}

\title{Image Classification in the Dark using \\ Quanta Image Sensors}
\author{Abhiram Gnanasambandam, and Stanley H. Chan,~\IEEEmembership{Senior Member,~IEEE}
\thanks{The authors are with the School of Electrical and Computer Engineering, Purdue University, West Lafayette, IN 47907, USA. Email: \texttt{\{agnanasa, stanchan\}@purdue.edu}. 

This work is supported, in part, by the National Science Foundation under grant CCF-1718007.

This paper is presented in the 16-th European Conference on Computer Vision (ECCV), Glasgow, United Kingdom, August 2020.}
}
\graphicspath{{./pix/}}

\IEEEtitleabstractindextext{\begin{abstract}
State-of-the-art image classifiers are trained and tested using well-illuminated images. These images are typically captured by CMOS image sensors with at least tens of photons per pixel. However, in dark environments when the photon flux is low, image classification becomes difficult because the measured signal is suppressed by noise. In this paper, we present a new low-light image classification solution using Quanta Image Sensors (QIS). QIS are a new type of image sensors that possess photon counting ability without compromising on pixel size and spatial resolution. Numerous studies over the past decade have demonstrated the feasibility of QIS for low-light imaging, but their usage for image classification has not been studied. This paper fills the gap by presenting a  student-teacher learning scheme which allows us to classify the noisy QIS raw data. We show that with student-teacher learning, we are able to achieve image classification at a photon level of one photon per pixel or lower. Experimental results verify the effectiveness of the proposed method compared to existing solutions.
\end{abstract}

}

\maketitle
\section{Introduction}

Quanta Image Sensors (QIS) are a type of single-photon image sensors originally proposed by E. Fossum as a candidate solution for the shrinking full-well capacity problem of the CMOS image sensors  (CIS) ~\cite{fossum2005gigapixel,fossum200611}. Compared to the CIS which accumulate photons to generate signals, QIS have a different design principle which partitions a pixel into many tiny cells called the jots with each jot being a single-photon detector. By oversampling the space and time, and by using a carefully designed image reconstruction algorithm, QIS can capture very low-light images with signal-to-noise ratio much higher than existing CMOS image sensors of the same pixel pitch \cite{Chan16}. Over the past few years, prototype QIS have been built by researchers at Dartmouth and Gigajot Technology Inc. \cite{ma2015pump,ma2017photon}, with a number of theoretical and algorithmic contributions by researchers at EPFL~\cite{yang2010optimal,chandramouli2019bit}, Harvard~\cite{chan2014efficient}, and Purdue \cite{Gnanasambandam:19,elgendy2019color,choi2018image,gnanasambandamhigh,elgendy2017optimal}. Today, the latest QIS prototype can perform color imaging with a read noise of 0.25$e^-$/pix (compared to at least several electrons in CIS\cite{fowler2013read})  and dark current of 0.068$e^-$/pix/s at room temperature (compared to $>1e^-$/pix/s in CIS) ~\cite{ma2017photon,Gnanasambandam:19}.

\begin{figure*}[!]
\centering
\includegraphics[width =  \linewidth]{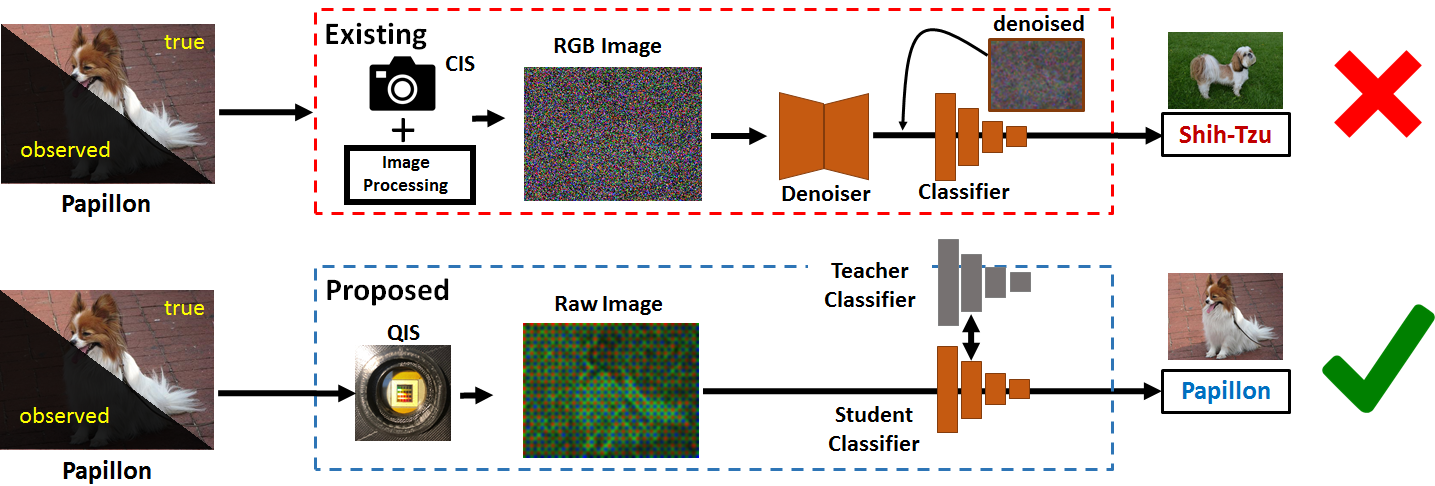}
\caption{\label{fig:Real_data_emph} [Top] Traditional image classification methods are based on CMOS image sensors (CIS), followed by a denoiser-classifier pipeline. [Bottom] The proposed classification method comprises a novel image sensor QIS and a novel student-teacher learning protocol. QIS generates significantly stronger signals, whereas student-teacher learning improves the robustness against noise.}
\end{figure*}

While prior work has demonstrated the effectiveness of using QIS for low-light image formation, there is no systematic study of how QIS can be utilized to perform better image classification in the dark. The goal of this paper is to fill the gap by proposing the first QIS image classification solution. Our proposed method is summarized in \fref{fig:Real_data_emph}. Compared to the traditional CIS-based low-light image classification framework, our solution leverages the unique single-photon sensing capability of QIS to acquire very low-light photon count images. We do not use any image processing, and directly feed the raw Bayer QIS data into our classifier. Our classifier is trained using a novel student-teacher learning protocol, which allows us to transfer knowledge from a teacher classifier to a student classifier. We show that the student-teacher protocol can effectively alleviate the need for a deep image denoiser as in the traditional frameworks. Our experiments demonstrate that the proposed method performs better than the existing solutions. The overall system -- QIS combined with student-teacher learning -- can achieve image classification on real data at 1 photon per pixel or lower. To summarize, the two contributions of this paper are:
\begin{enumerate}
\item[(i)] The introduction of student-teaching learning for low-light image classification problems. The experiments show that the proposed method outperforms existing approaches.
\item[(ii)] The first demonstration of image classification at a photon level of 1 photon per pixel or lower, on real images. This is a very low photon level compared to other results reported in the image classification literature.
\end{enumerate}

\begin{figure*}[!]
\centering
    \begin{tabular}{cccc}
    \hspace{-2ex}\includegraphics[width=0.24\linewidth]{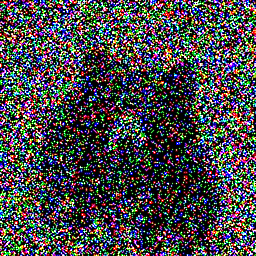}&
    \hspace{-2ex}\includegraphics[width=0.24\linewidth]{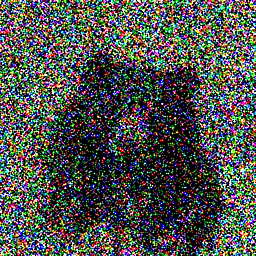}&
    \hspace{-2ex}\includegraphics[width=0.24\linewidth]{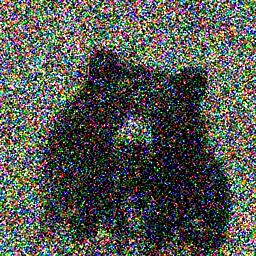}&
    \hspace{-2ex}\includegraphics[width=0.24\linewidth]{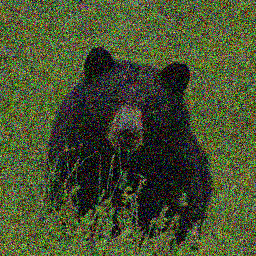}\\
    QIS 0.25 ppp & QIS 1 ppp & QIS 4 ppp & i.i.d. Gaussian \\ & & & $\sigma = 100/255$
    \end{tabular}
    \vspace{-2ex}
    \caption{\textbf{How Dark is One Photon Per Pixel?} The first three sub-images in this figure are the real captures by a prototype QIS at various photon levels. The last sub-image is a simulation using additive i.i.d. Gaussian noise of a level of $\sigma = 100/255$, which is often considered as heavy degradation in the denoising literature. Additional examples can be found in \fref{fig:Real_data}.}
    \vspace{-2ex}
\label{fig:Noisy}
\end{figure*}

\section{Background}
\subsection{Quanta Image Sensor}
Quanta Image Sensors are considered as one of the most promising candidates for the third generation image sensors after CCD and CMOS. Fabricated using the commercial 3D stacking technology, the current sensor has a pixel pitch of 1.1$\mu$m, with even smaller sensors being developed. The advantage of QIS over the conventional CMOS image sensors is that at 1.1$\mu$m, the read noise of QIS is as low as $0.25e^-$ whereas a typical 1$\mu$m CMOS image sensor is at least several electrons. This low read noise (and also the low dark current) is made possible by the unique non-avalanche design \cite{ma2017photon} so that pixels can be packed together without causing strong stray capacitance. The non-avalanche design also differentiates QIS from single photon avalanche diodes (SPAD). SPADs are typically bigger in size $>5\mu$m, have lower fill factor $<70\%$, have lower quantum efficiency $<50\%$, and have significantly higher dark count $> 10e^-$. See \cite{Gnanasambandam:19} for a detailed comparison between CIS, SPAD and QIS. In general, SPADs are useful for applications such as time-of-flight imaging because of their speed \cite{o2017reconstructing,lindell2018single,callenberg2019emccd,gariepy2015single}, although new results in HDR imaging has been reported \cite{Gupta_ICCV2019,Ma_SIGGRAPH20}. QIS have better resolution and works well for passive imaging.

\subsection{How Dark is One Photon Per Pixel?}
When we say low-light imaging, it is important to clarify the photon level. The photon level is usually measured in the terms of lux. However, a more precise definition is the unit of photons per pixel (ppp). ``Photons per pixel'' is the average number of photons a pixel sees during the exposure period. We use photons per pixel as the metric because the amount of photons detected by a sensor depends on the exposure time and sensor size --- A large sensor inherently detects more photons, so does long exposure. For example, under the same low-light condition, images formed by the Keck telescope (aperture diameter = 10m) certainly has better signal-to-noise than an iPhone camera (aperture diameter = 4.5mm). A high-end 3.5$\mu$m camera today has a read noise greater than $2e^-$ \cite{Sony_rn}. Thus, our benchmark choice of 1 ppp is approximately half of the read noise of a high-end sensor today. To give readers an idea of the amount of noise we should expect to see at 1 ppp, we show a set of real QIS images in \fref{fig:Noisy}. Signals at 1 ppp is significantly worse than the so called ``heavy noise'' images we observe in the denoising literature and the low-light classification literature.

\begin{figure*}[!]
    \centering
    \includegraphics[width=\linewidth]{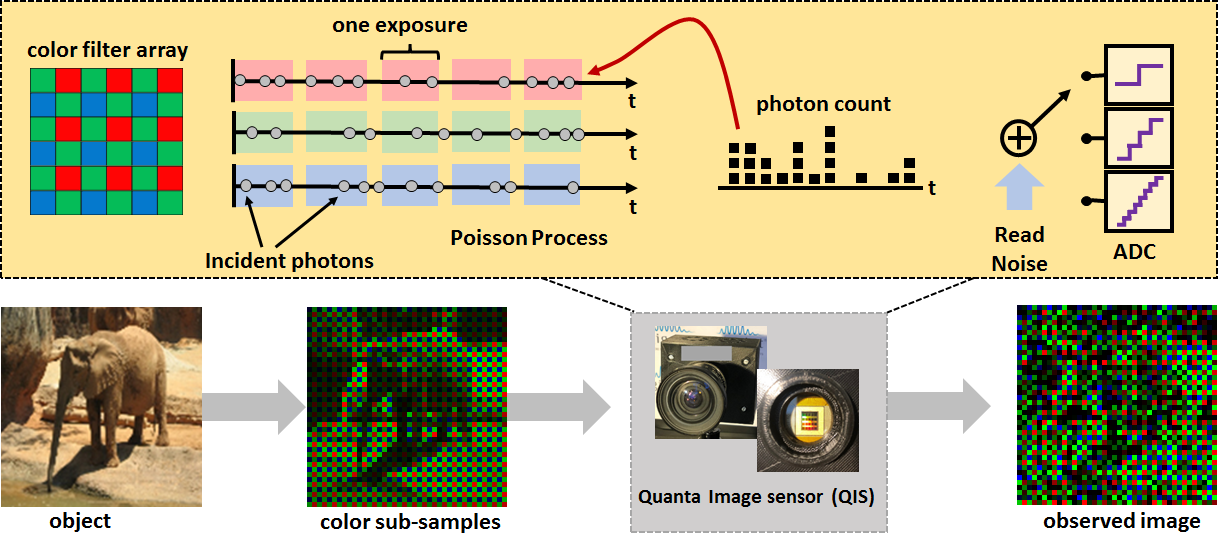}
    \vspace{-4ex}
    \caption{\textbf{QIS Image Formation Model}. The basic image formation of QIS consists of a color filter array, a Poisson process, read noise, and an analog-to-digital converter (ADC). Additional factors are summarized in \eref{eq: QIS equation}. }
    \label{fig:image_formation}
    \vspace{-2ex}
\end{figure*}

\subsection{Prior Work}
\noindent\textbf{Quanta Image Sensors}. QIS were proposed in 2005, and since then significant progresses have been made over the past 15 years. Readers interested in the sensor development can consult recent keynote reports, e.g., \cite{fossum2016quanta}. On the algorithmic side, a number of theoretical signal processing results and reconstruction algorithms have been proposed \cite{Chan16, chan2016plug,Gnanasambandam:19}, including some very recent methods based on deep learning \cite{choi2018image,chandramouli2019bit}. However, since the sensor is relatively new, computer vision applications of the sensor are not yet common. To the best of our knowledge, the only available method for tracking applications is \cite{gyongy2018single}.

\vspace{1ex}
\noindent\textbf{Low-light Classification}. Majority of the existing work in classification is based on well-illuminated CMOS images. The first systematic study of the feasibility of low-light classification was presented by Chen and Perona \cite{chen2017seeing}, who observed that low-light classification is achievable by using a few photons. In the same year, Diamond et al. \cite{diamond2017dirty} proposed the ``Dirty Pixels'' method by training a denoiser and a classifier simultaneously. They observed that less aggressive denoisers are better for classification because the features are preserved. Other methods adopt similar strategies, e.g., using discrete cosine transform \cite{hossain2019distortion}, training a classifier to help denoising \cite{wu2017relation} or using an ensemble method \cite{dodge2017quality}, or training a denoiser that are better suited for pre-trained classifiers \cite{liu2018connecting,liu2019transferable}.

\vspace{1ex}
\noindent\textbf{Low-light Reconstruction}.
A closely related area of low-light classification is low-light reconstruction, e.g., denoising. Classical low-light reconstruction usually follows the line of Poisson-based inverse problems \cite{makitalo2010optimal} and contrast enhancement \cite{malm2007adaptive,hu2014deblurring,guo2016lime, fu2018retinex}. Deep neural network methods have recently become the main driving force \cite{remez2017deep, zhang2017beyond,lore2017llnet,zhang2018ffdnet, plotz2017benchmarking, xu2018real}, including the recent series on ``seeing in the dark'' by Chen et al. \cite{chen2018learning,chen2019seeing}. Burst photography \cite{hasinoff2016burst,mildenhall2018burst,davy2018non,kokkinos2019iterative} (with some older work in  \cite{liu2010high,liu2014fast,joshi2010seeing}) is related but not directly applicable to us since the methods are developed for multi-frame problems.

\section{Method}
The proposed method comprises QIS and a novel student-teacher learning scheme. In this section, we first discuss how images are formed by QIS. We will then present the proposed student-teacher learning scheme which allows us to overcome the noise in QIS measurements.

\subsection{QIS Image Formation Model} \label{sec:Imageformation}
  The image formation model is shown in \fref{fig:image_formation}. Given an object in the scene ($\vx_\text{rgb}$), we use a color filter array (CFA) to bandpass the light to subsample the color. Depending on the exposure time and the size of the jots, a sensor gain $\alpha$ is applied to scale the sub-sampled color pixels. The photon arrival is simulated using a Poisson model. Gaussian noise is added to simulate the read noise arising from the circuit. Finally, an analog-to-digital converter (ADC) is used to truncate the real numbers to integers depending on the number of bits allocated by the sensor. For example, a single-bit QIS will output two levels, whereas multi-bit QIS will output several levels. In either case, the signal is clipped to take value in $\{0,1,\hdots L\}$, where $L$ represents the maximum signal level. The image formation process can be summarized using the following equation
\begin{align}
\underset{\R^{M \times N}}{ \underbrace{\vx_\text{QIS}}}
&= \text{ADC}_{[0,L]}\bigg\{ \underset{
\text{photon arrival}}{\underbrace{\text{Poisson}}}\bigg( \underset{\text{sensor gain}}{\underbrace{\alpha}} \label{eq: QIS equation} \\
&\qquad\qquad\qquad \;\times\; \text{CFA}\big( \underset{\R^{M\times N \times 3}}{ \underbrace{\vx_{\text{rgb}}}}\big)\bigg) +
\underset{\text{read noise}}{\underbrace{\veta}}\bigg\},\notag
\end{align}
In addition to the basic image formation model described in \eref{eq: QIS equation}, two other components are included in the simulations. First, we include the dark current which is an additive noise term to $\alpha \cdot \text{CFA}(\vx_{\text{rgb}})$. The typical dark current of the QIS is $0.068e^-$/pix/s. Second, we model the pixel response non-uniformity (PRNU). PRNU is a pixel-wise multiplication applied to $\vx_{\text{rgb}}$, and is unique for every sensor. Readers interested in details on the image formation model and statistics can consult previous works such as \cite{yang2011bits,fossum2013modeling,Chan16,elgendy2017optimal}.

\begin{figure*}[!]
\centering
\includegraphics[width = 0.8\linewidth]{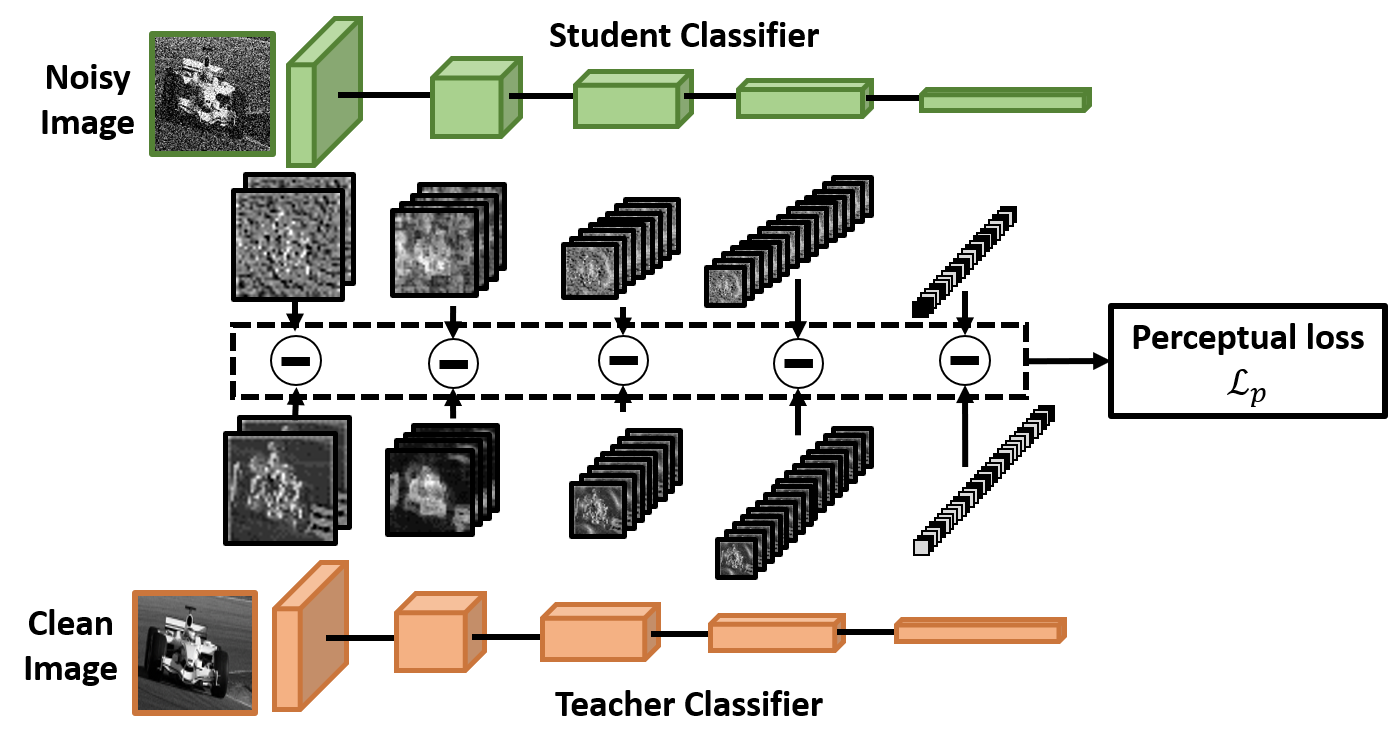}
\vspace{-2ex}
\caption{\textbf{Student-Teacher Learning}. Student-teacher learning comprises two networks: A teacher network and a student network. The teacher network is pre-trained using clean samples whereas the student is trained using noisy samples. To transfer knowledge from the teacher to the student, we compare the features extracted by the teacher and the student at different stages of the network. The difference between the features is measured as the perceptual loss.}
\vspace{-2ex}
\label{fig: student teacher}
\end{figure*}

\begin{figure*}[!]
\centering
    \begin{tabular}{cc}
    \includegraphics[width=0.46\linewidth]{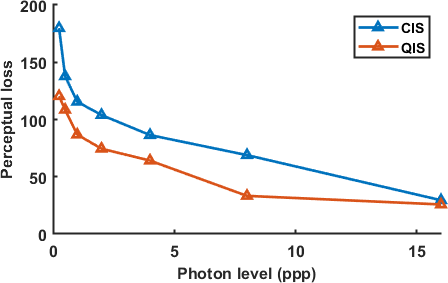} &
    \includegraphics[width=0.46\linewidth]{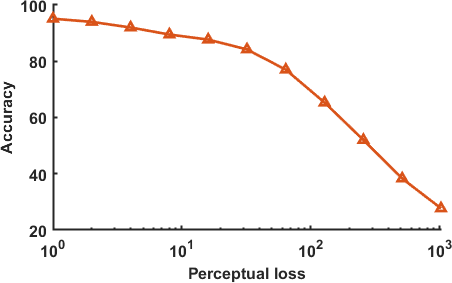}\\
    (a) Perceptual loss vs Photon level & (b) Accuracy vs Perceptual loss
    \end{tabular}
\caption{\textbf{Effectiveness of Student-Teacher Learning}. (a) Perceptual loss as a function of photon level. (b) Classification accuracy as a function of the perceptual loss $\calL_p(\vx_{\text{QIS}}, \vx_{\text{rgb}})$. The accuracy is measured by repeating the synthetic experiment described in the Experiment Section. The negative correlation suggests that perceptual loss is indeed an influential factor. }
\label{fig: benefit}
\vspace{-2ex}
\end{figure*}

\subsection{Student-Teacher Learning}
\label{sec: student teacher}
Inspecting \eref{eq: QIS equation}, we notice that even if the read noise $\veta$ is zero, the random Poisson process will still create a fundamental limit due to the shot noise in $\vx_{\text{QIS}}$. Therefore, when applying a classification method to the raw QIS data, some capability of removing the shot noise becomes necessary. Traditional solution to this problem (in the context of CIS) is to denoise the images as shown in the top of \fref{fig:Real_data_emph}. The objective of this section is to introduce an alternative approach using the concept of student-teacher learning.

The idea of student-teacher learning can be understood from \fref{fig: student teacher}. There are two networks in this figure: A teacher network and a student network. The teacher network is trained using \emph{clean} samples, and is pre-trained, i.e., its network parameters are fixed during training of the student network. The student network is trained using \emph{noisy} samples with the assistance from the teacher. Because the teacher is trained using clean samples, the features extracted are in principle ``good'', in contrast to the features of the student which are likely to be ``corrupted''. Therefore, in order to transfer knowledge from the teacher to the student, we propose minimizing a \emph{perceptual loss} as defined below. We define the $j$-th layer's feature of the student network as $\phi^j(\vx_{\text{QIS}})$, where $\phi^j(\cdot)$ maps $\vx_{\text{QIS}}$ to a feature vector, and we define $\widehat{\phi}^j(\vx_{\text{rgb}})$ as the feature vector extracted by the teacher network. The perceptual loss is
 \begin{equation}
    \calL_{\text{p}}(\vx_{\text{QIS}}, \vx_{\text{rgb}}) =
    \sum_{j=1}^J \underset{\text{$j$-th layer's perceptual loss}}{\underbrace{\frac{1}{N_j}\left\|\widehat{\phi}^j(\vx_{\text{rgb}}) - \phi^j(\vx_{\text{QIS}}) \right\|^2}},
\end{equation}
where $N_j$ is the dimension of the $j$-th feature vector. Since the perceptual loss measures the distance between the student and the teacher, minimizing the perceptual loss forces them to be close. This, in turn, forces the network to ``denoise'' the shot noise and read noise in $\vx_{\text{QIS}}$ before predicting the label.

\begin{figure*}[!]
    \centering
    \includegraphics[width = 0.9\linewidth]{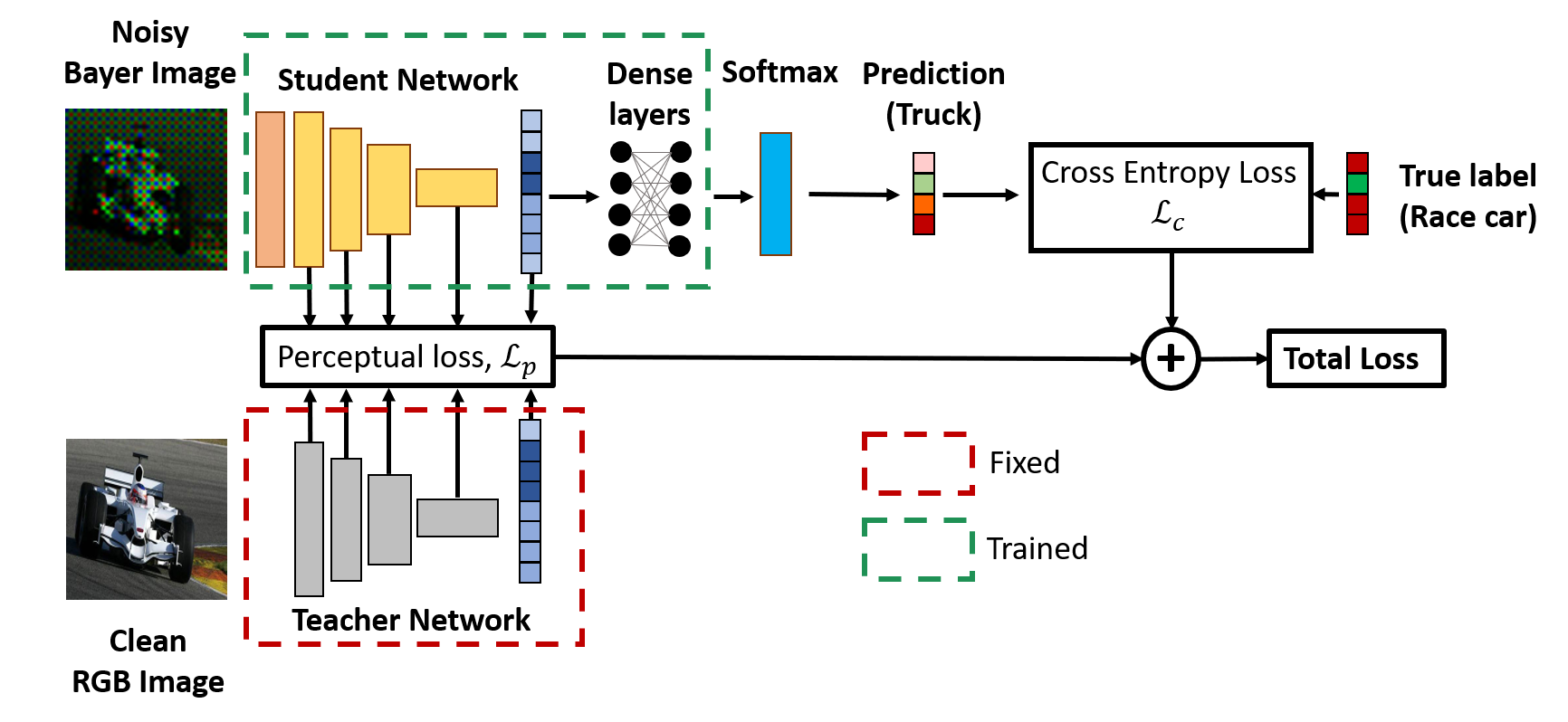}
    \vspace{-2ex}
    \caption{\textbf{Proposed Method}. The proposed method trains a classification network with two training losses: (1) cross-entropy loss to measure the prediction quality, and (2) perceptual loss to transfer knowledge from teacher to student. During testing, only the student is used. We introduce a 2-layer entrance (colored in orange) for the student network so that the classifier can handle the Bayer image.}
    \label{fig:JPLNet}
    \vspace{-2ex}
\end{figure*}

We conduct a simple experiment to demonstrate the impact of input noise to perceptual loss and  classification accuracy. We first consider a pre-trained teacher network by sending QIS data at different photon levels. As photon level drops, the quality of the features also drops and hence the perceptual loss increases. This is illustrated in \fref{fig: benefit}(a). Then in \fref{fig: benefit}(b), we evaluate the classification accuracy by using the synthetic testing data outlined in the Experiment Section. As the perceptual loss increases, the classification accuracy drops. This result suggests that if we minimize the perceptual loss then the classification accuracy can be improved.

Our proposed student-teacher learning is inspired by the knowledge distillation work of Hinton et al. \cite{hinton2015distilling} which proposed an effective way to compress networks. A number of follow up ideas have been proposed, e.g., \cite{ba2014deep, zhang2018deep, yang2019training, guo2019robust}, including the MobileNet \cite{howard2017mobilenets}. The concept of perceptual loss has been used in various computer vision applications such as the texture-synthesis and style-transfer by Johnson et al. \cite{johnson2016perceptual} and Gatys et al. \cite{gatys2015texture, gatys2015neural}, among many others \cite{chen2016joint, liu2018connecting, talebi2018learned, mahendran2015understanding, simonyan2013deep,yosinski2015understanding,ma2015pump}. The method we propose here is different in the sense that we are not compressing the network. Also, we are not asking the student to mimic the teacher because the teacher and the student are performing two different tasks: The teacher classifies clean data, whereas the student classifies noisy data.  In the context of low-light classification, student-teacher learning has not been applied.

\subsection{Overall Method}
The overall loss function comprises the perceptual loss and the conventional prediction loss using cross-entropy. The cross-entropy loss $\calL_{\text{c}}$, measures the difference between true label $y$ and the predicted label $f_{\mTheta}(\vx_{\text{QIS}})$ generated by the student network, where $f_{\mTheta}$ is the student network. The overall loss is mathematically described as
\begin{align}\label{eqn:comb_perc}
\calL(\mTheta) = \sum_{n=1}^N \bigg\{\calL_{\text{c}}\big(y^n, f_{\mTheta}(\vx_{\text{QIS}}^n)\big) + \lambda \calL_{\text{p}}\big(\vx_{\text{rgb}}^n,  \vx_{\text{QIS}}^n \big)\bigg\},
\end{align}
where $\vx^n$ denotes the $n$-th training sample with the ground truth label $y^n$. During the training, we optimize the weights of the student network by solving
\begin{equation}
    \widehat{\mTheta} = \argmin{\mTheta} \; \calL(\mTheta).
\end{equation}
During testing, we feed a testing sample $\vx_{\text{QIS}}$ to the student network and evaluate the output:
\begin{equation}
    \widehat{y} = f_{\widehat{\mTheta}}(\vx_{\text{QIS}}).
\end{equation}

\fref{fig:JPLNet} illustrates the overall network architecture. In this figure, we emphasize that training is done on the student only. The teacher is fixed and is not trainable. In this particular example, we introduce a very shallow network consisting of 2 convolution layers with 32 and 3 filters respectively. This shallow network is used to perform the necessary demosaicking by converting raw Bayer pattern to the full RGB before feeding into a standard classification network.

\section{Experiments}
\subsection{Dataset}
\textbf{Dataset}. We consider two datasets. The first dataset (Animal) contains visually distinctive images where the class labels are far apart. The second dataset (Dog) contains visually similar images where the class labels are fine-grained. The two different datasets can help differentiating the performance regime of the proposed method, and its benefits over other state-of-the-art networks.

\begin{figure*}[!]
\centering
    \begin{tabular}{cc}
    \includegraphics[width=0.48\linewidth]{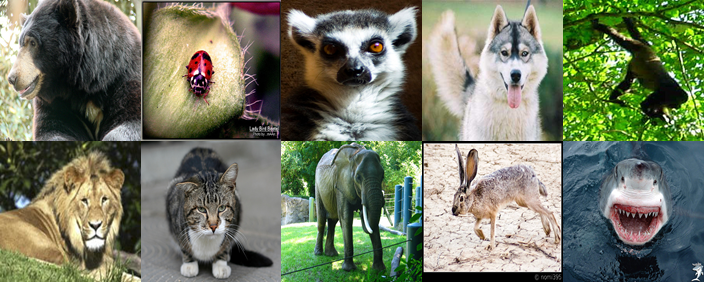} &
        \includegraphics[width=0.48\linewidth]{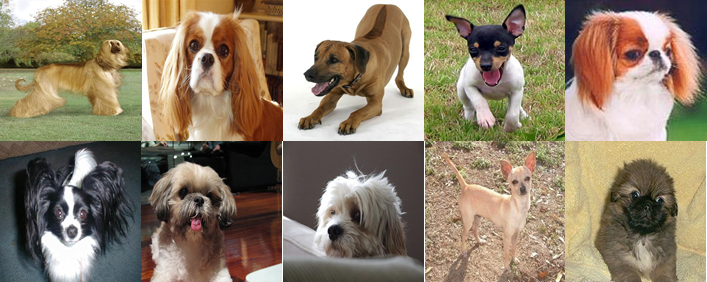} \\
    (a) Animal Dataset &  (b) Dog Dataset
    \end{tabular}
    \caption{The two datasets for our experiments. }
\label{fig:snapshot}
\end{figure*}

\begin{figure*}[!]
\centering
    \begin{tabular}{cc}
    \includegraphics[width=0.45\linewidth]{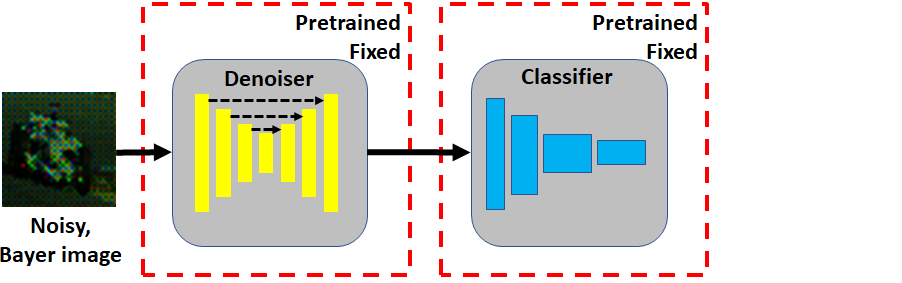} & 
    \includegraphics[width=0.45\linewidth]{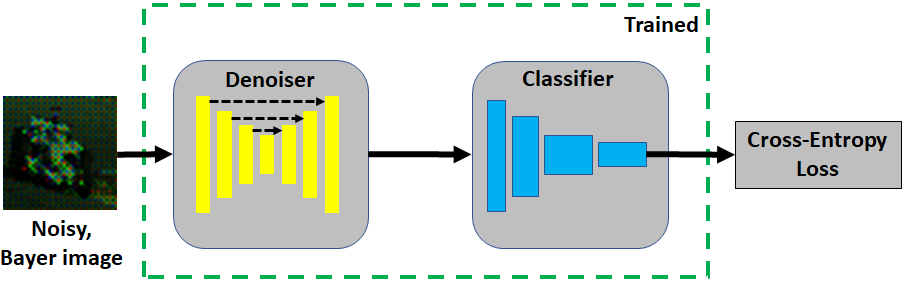} \\
    (a) Vanilla Network & (b) Dirty Pixels \cite{diamond2017dirty} \\
    \includegraphics[width=0.45\linewidth]{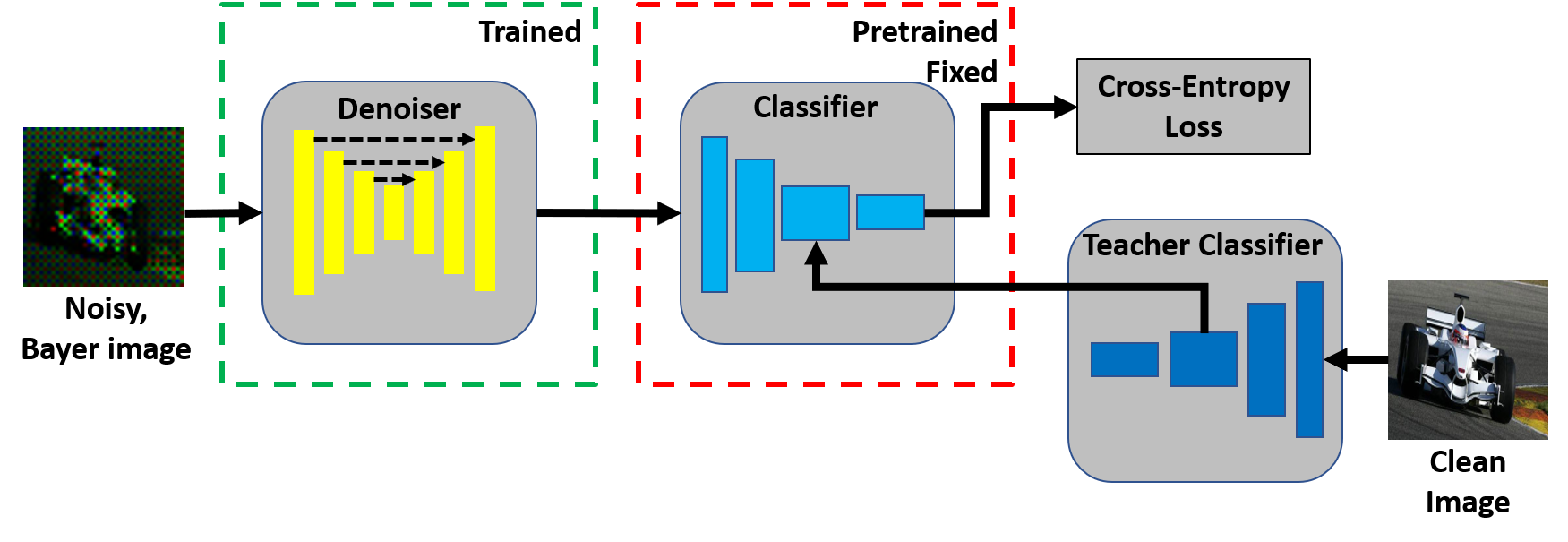} &
    \includegraphics[width=0.45\linewidth]{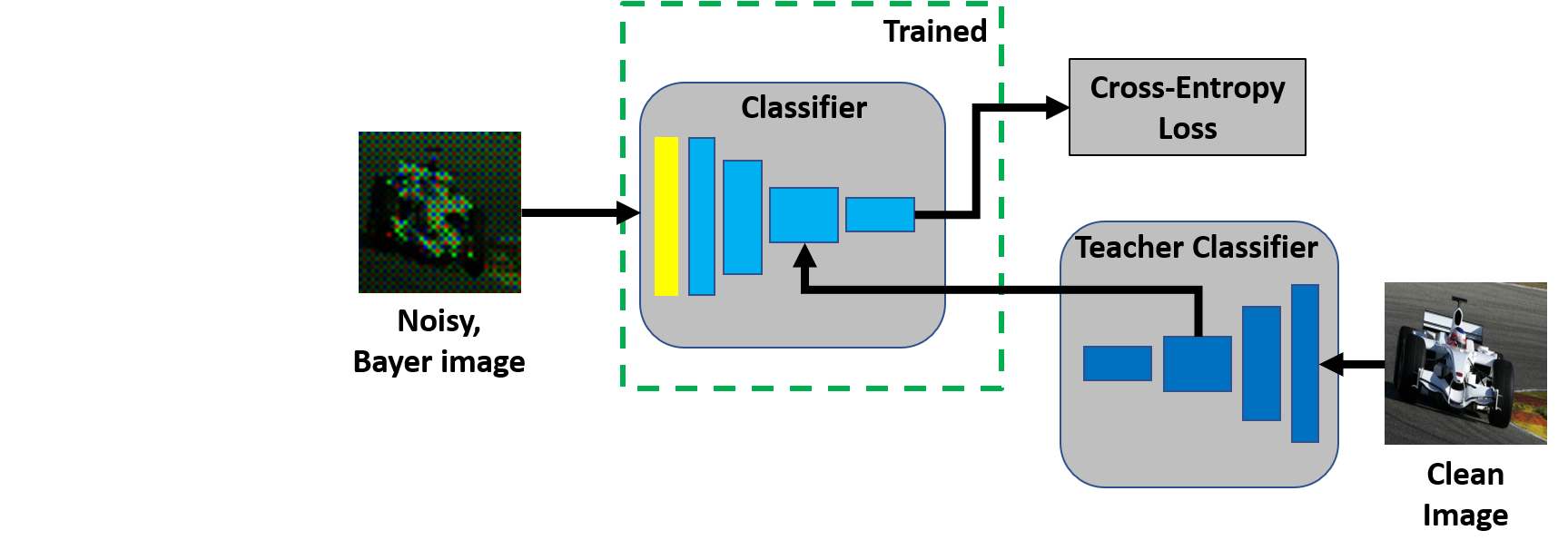} \\
    (c) Restoration Network \cite{liu2018connecting} & (d) Proposed Method  \\
    \end{tabular}
    \caption{\textbf{Competing Methods}. The major difference between the networks are the trainable modules and the loss functions. For Dirty Pixels and our proposed method, we further split it into two versions: Using a deep denoiser or using a shallow entrance network.}
\label{fig:methods_comparison}
\end{figure*}

The construction of the two datasets are as follows. For the Animal dataset, we randomly select 10 classes of animals from ImageNet \cite{deng2009imagenet}, as shown in \fref{fig:snapshot}(a). Each class contains 1300 images, giving a total of 13K images. Among which, 9K are used for training, 1K for validation, and 3K for testing. For the Dog dataset, we randomly select 10 classes of dogs from the Stanford Dog dataset \cite{khosla2011novel}, as shown in \fref{fig:snapshot}(b). Each class has approximately 150 images, giving a total of 1919 images. We use 1148 for training, 292 for validation, and 479 for testing.

\vspace{-2ex}
\subsection{Competing Methods and Our Network}
We compare our method with three existing low-light classification methods as shown in \fref{fig:methods_comparison}. The three competing methods are (a) Vanilla denoiser + classifier, an ``off-the-shelf'' solution using pre-trained models but fine-tuned using QIS data. THe denoiser is pre-trained on the QIS data and the classifier is pre-trained on clean images. (b) Dirty Pixels \cite{diamond2017dirty}, same as Vanilla denoiser + classifier, but trained end-to-end using the QIS noisy data. (c) Restoration Network \cite{liu2018connecting,liu2019transferable}, which trains a denoiser but uses a pre-trained classifier. This can be viewed as a middle-ground solution between Vanilla and Dirty Pixels.

To ensure that the comparison is fair w.r.t. the training protocol and not the architecture, all classifiers in this experiment (including ours) use the same VGG-16 architecture. For methods that use a denoiser, the denoiser is fixed as a UNet. This particular combination of denoiser and classifier does have some influence to the final performance, but the effectiveness of the training protocol can still be observed. Combinations beyond the ones we report here can be found in the ablation study. For Dirty Pixels and our proposed method, we further split them into two versions: (i) Using a deep denoiser as the entrance, i.e., a 20-layer UNet, and (ii) using a shallow two-layer network as the entrance to handle the Bayer pattern, as we described in the proposed method section. We will analyze the influence of this component in the ablation study.

\begin{figure*}[!]
\centering
    \begin{tabular}{cc}
    \includegraphics[width=0.5\linewidth]{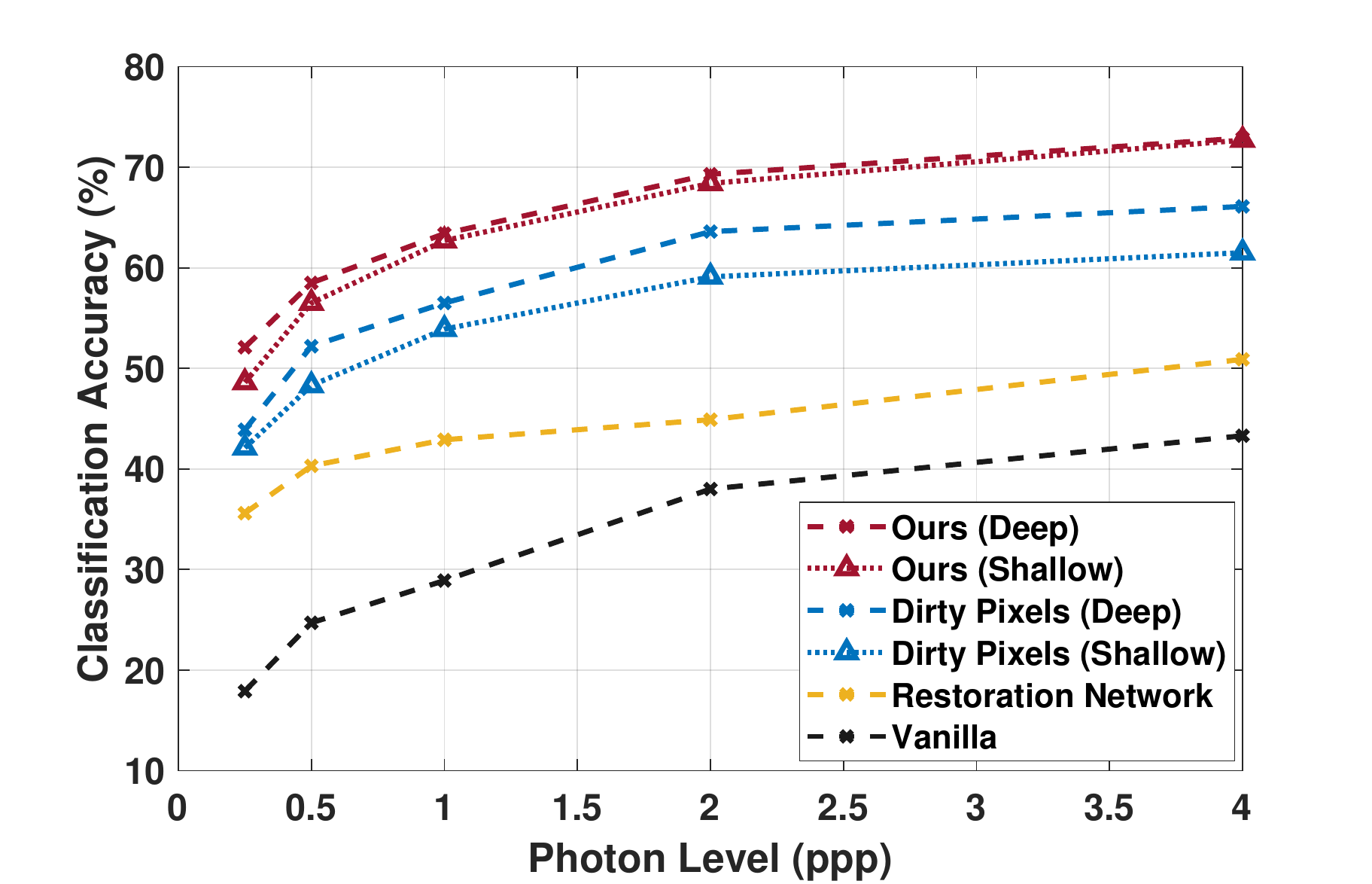} &
       \hspace{-4ex}
       \includegraphics[width=0.5\linewidth]{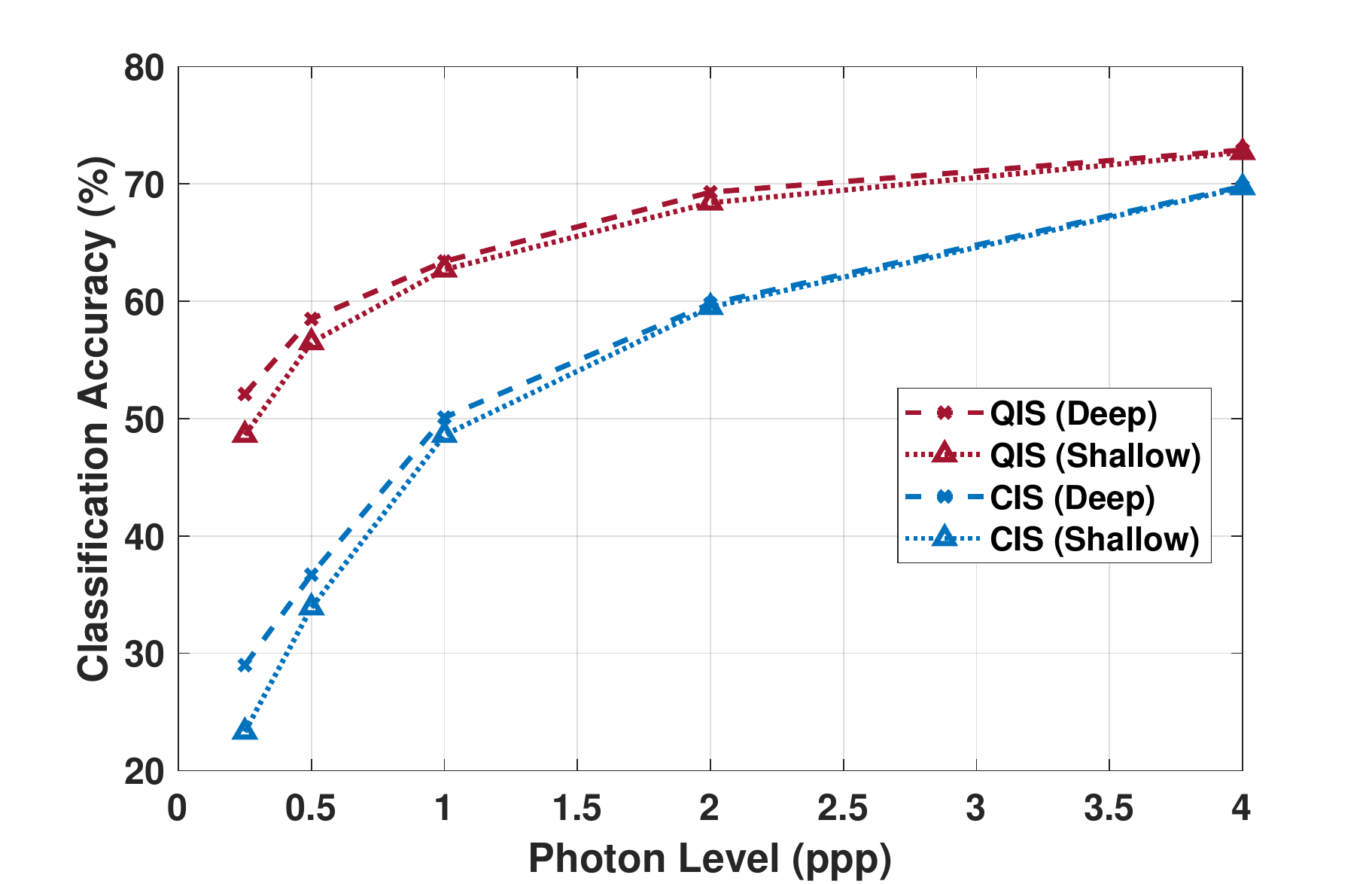} \\
    (a) Different Classifiers &  (b) Different Sensors
    \end{tabular}
    \vspace{-1ex}
    \caption{\textbf{Synthetic Data on Dog Dataset}. (a) Comparing different classification methods using QIS as the sensor. (b) Comparing QIS and CIS using our proposed classifier. }
\label{fig:results}
\end{figure*}

\subsection{Synthetic Experiment}
The first experiment is based on synthetic data. The training data are created by the QIS model. To simulate the QIS data, we follow Equation \eref{eq: QIS equation} by using the Poisson-Gaussian process.   the read noise is $\sigma = 0.25e^-$ according to \cite{ma2017photon}. The analog-to-digital converter is set to 5 bits, so that the number of photons seen by the sensors is between 0 and 31. We use a similar simulation procedure for CIS with the difference being the read noise, which we set to $\sigma = 2.0e^-$ \cite{Sony_rn}.

The experiments are conducted for 5 different photon levels corresponding to 0.25, 0.5, 1, 2, and 4 photons per pixel (ppp). The photon level is controlled by adjusting the value of the multiplier $\alpha$ in Equation \eref{eq: QIS equation}. The loss function weights $\lambda$ in Equation \eref{eqn:comb_perc} is tuned for optimal performance.

The results of the synthetic data experiment are shown in \fref{fig:results}. In \fref{fig:results}(a), we observe that our proposed classification is consistently better than competing methods the photon levels we tested. Moreover, since all methods reported in \fref{fig:results}(a) are using QIS as the sensor, the curves in \fref{fig:results}(a) reveal the effectiveness of just the classification method. In \fref{fig:results}(b), we compare the difference between using QIS and CIS. As we expect, CIS has worse performance compared to QIS.

\begin{figure*}[!]
\centering
    \begin{tabular}{ccccccc}
    & Ground Truth &QIS &CIS & Ground Truth &QIS &CIS  \\
    \rotatebox{90}{\ \ \ \ \ \ \ \ \ \ \ \ \ \textbf{0.25 $e^-$}} &
    \hspace{-2ex}\includegraphics[width=0.15\linewidth]{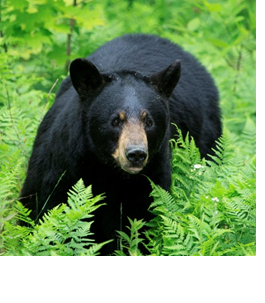}&
    \hspace{-2ex}\includegraphics[width=0.15\linewidth]{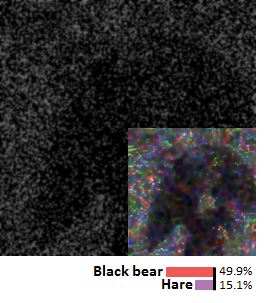}&
    \hspace{-2ex}\includegraphics[width=0.15\linewidth]{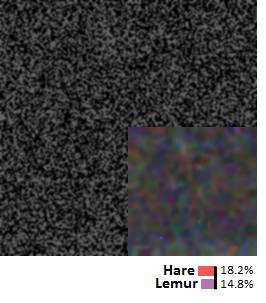}&
    \hspace{-2ex}\includegraphics[width=0.15\linewidth]{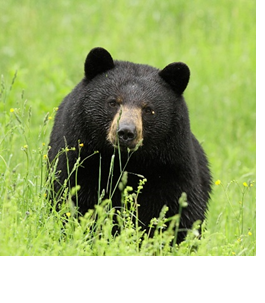}&
    \hspace{-2ex}\includegraphics[width=0.15\linewidth]{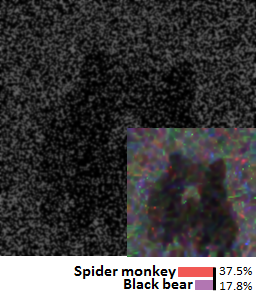}&
    \hspace{-2ex}\includegraphics[width=0.15\linewidth]{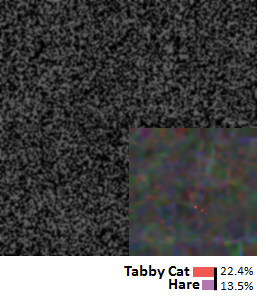}\\
    \rotatebox{90}{\ \ \ \ \ \ \ \ \ \ \ \ \ \textbf{0.50 $e^-$}} &
    \hspace{-2ex}\includegraphics[width=0.15\linewidth]{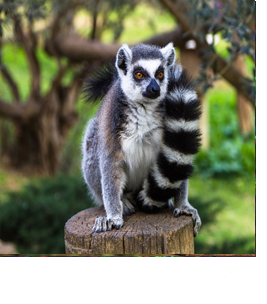}&
    \hspace{-2ex}\includegraphics[width=0.15\linewidth]{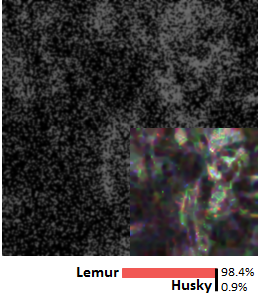}&
    \hspace{-2ex}\includegraphics[width=0.15\linewidth]{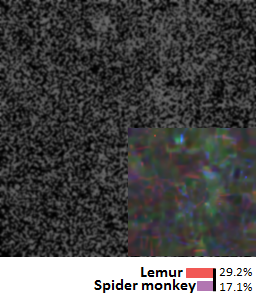}&
    \hspace{-2ex}\includegraphics[width=0.15\linewidth]{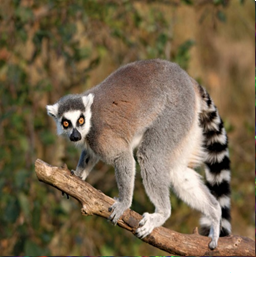}&
    \hspace{-2ex}\includegraphics[width=0.15\linewidth]{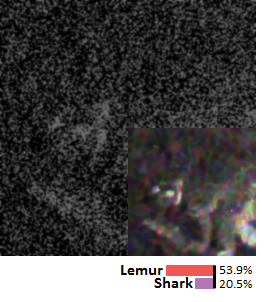}&
    \hspace{-2ex}\includegraphics[width=0.15\linewidth]{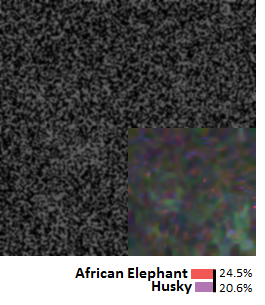}\\
    \rotatebox{90}{\ \ \ \ \ \ \ \ \ \ \ \ \ \textbf{1.00 $e^-$}} &
    \hspace{-2ex}\includegraphics[width=0.15\linewidth]{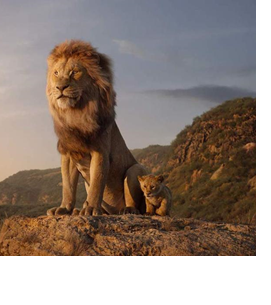}&
    \hspace{-2ex}\includegraphics[width=0.15\linewidth]{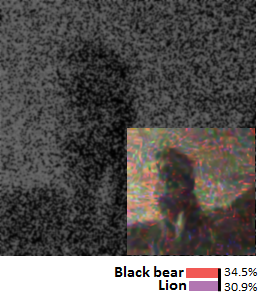}&
    \hspace{-2ex}\includegraphics[width=0.15\linewidth]{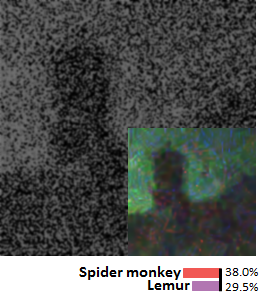}&
    \hspace{-2ex}\includegraphics[width=0.15\linewidth]{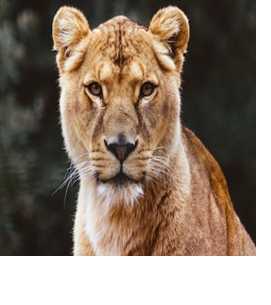}&
    \hspace{-2ex}\includegraphics[width=0.15\linewidth]{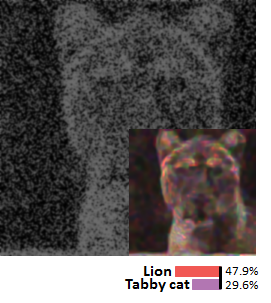}&
    \hspace{-2ex}\includegraphics[width=0.15\linewidth]{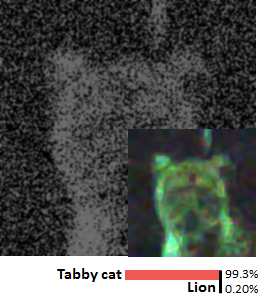}\\
    \end{tabular}
    \caption{\textbf{Real Image Results}. This figure shows raw Bayer data obtained from a prototype QIS and a commercially available CIS, and how they are classified using our proposed classifier. The inset images show the denoised images (by \cite{chen2018learning}) for visualization. Notice the heavy noise at $0.25$ and $0.5$ ppp, only QIS plus our proposed classification method can produce the correct prediction.}
\label{fig:Real_data}
\end{figure*}

\begin{figure*}[!]
\centering
    \begin{tabular}{cc}
    \includegraphics[width=0.48\linewidth]{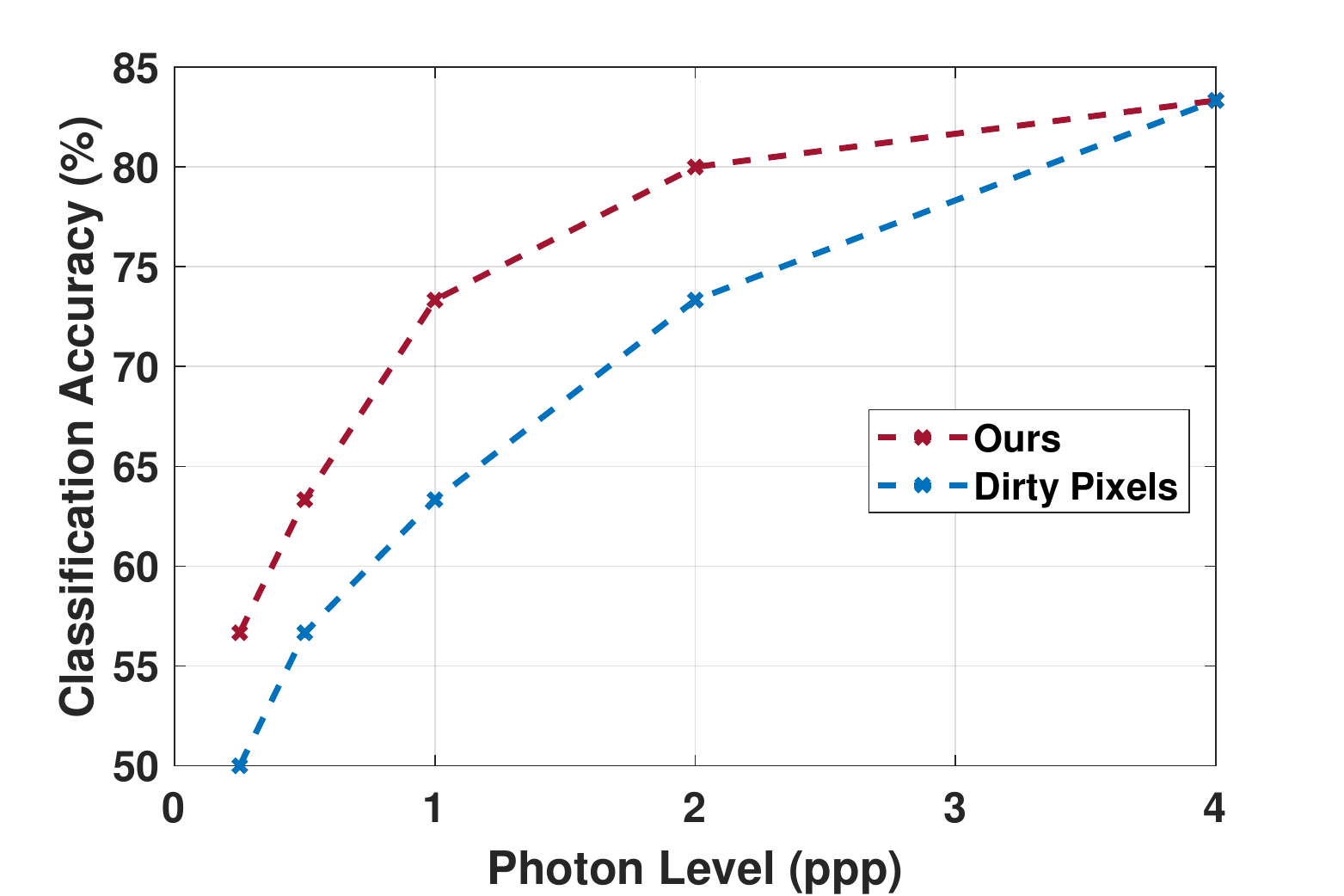} &
       \hspace{1.0ex} \includegraphics[width=0.48\linewidth]{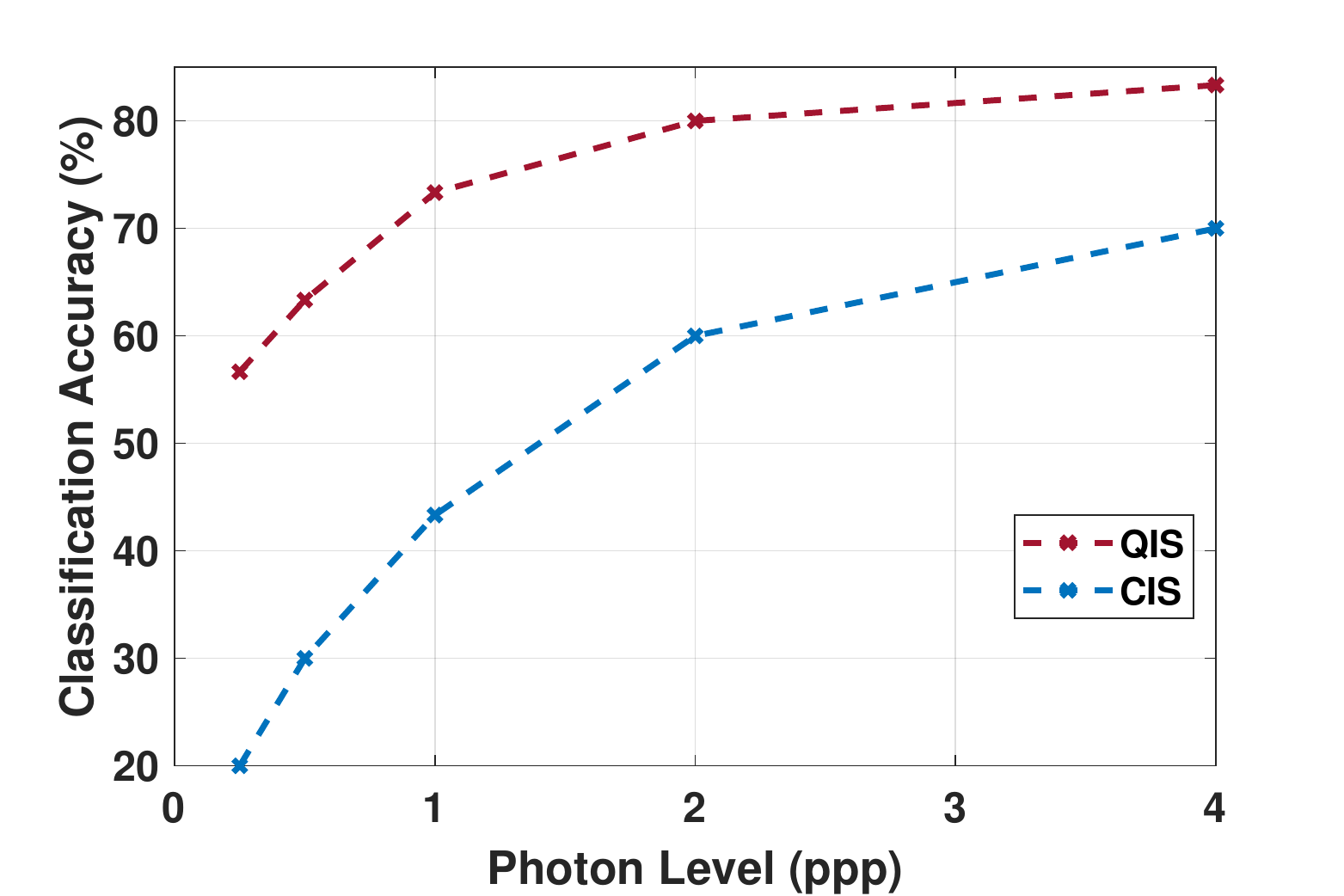} \\
    (a) Different Classifiers &\hspace{1.0ex}  (b) Different Sensors
    \end{tabular}
    \caption{\textbf{Real data on Animal Dataset}. (a) Comparing different classification methods using QIS as the sensor. (b) Comparing QIS and CIS using our proposed classifier. }
    \vspace{-3ex}
\label{fig:real_results}
\end{figure*}

\subsection{Real experiment}
We conduct an experiment using real QIS and CIS data. The real QIS data are collected by
a prototype QIS camera Gigajot PathFinder \cite{Gnanasambandam:19}, whereas the real CIS data are collected by using a commercially available camera. To setup the experiment, we display the images on a Dell P2314H LED screen (60Hz). The cameras are positioned 1m from the display so that the field of view covers $256 \times 256$ pixels of the image. The integration time of the CIS is set to 250$\mu$s and that of QIS is $75 \mu$s. Since the CIS and QIS have different lenses, we control their aperture sizes and the brightness of the screen such that the average number of photons per pixel is equal for both sensors.

The training of the network in this real experiment is still done using the synthetic dataset, with the image formation model parameters matched with the actual sensor parameters. However, since the real image sensors have pixel non-uniformity, during the training we multiply a random PRNU mask to each of the generated images in order to mimic the process of PRNU. For testing, we collect 30 real images at each photon level, across 5 different photon levels. This corresponds to a total of 150 real testing images.

When testing, we make two pairs of comparisons: Proposed (shallow) versus Dirty Pixels (shallow), and QIS versus CIS. The result of the first experiment is shown in \fref{fig:real_results}(a), where we observe that the proposed method has a consistent improvement over Dirty Pixels. The comparison between QIS and CIS is shown in \fref{fig:real_results}(b). It is evident that QIS has better performance compared to CIS. \fref{fig:Real_data} shows the visualizations. The ground truth images were displayed on the screen, and the background images in QIS and CIS column are actual measurements from the corresponding cameras, cropped to $256\times256$. The thumbnail images in the front are the denoised images for reference. They are not used during the actual classification. The color bars at the bottom report the confidence level of the predicted class. Note the significant visual difference between QIS and CIS, and the classification results.

\subsection{Ablation Study}
\label{sec: ablation}
In this section we report several ablation study results, and highlight the most influencing factors to the design.

\vspace{1.0ex}
\noindent
\textbf{Sensor}. Our first ablation study is to fix the classifier but change the sensor from QIS to CIS. This experiment will underline the impact of the sensor in the overall pipeline. The result of this ablation study can be seen from \fref{fig:results}(b). Specifically, at 4 ppp (high photon level) of the Dogs dataset, QIS + proposed has a classification accuracy of 72.9\% while CIS has 69.8\%. The difference is 3.1\%. As the photon level drops, the gap between QIS and CIS this gap widens to 23.1\% at 0.25 ppp. A similar trend is found in the Animals dataset. Thus at low light QIS has a clear advantage, although CIS can catch up when there is sufficient number of photons.

\vspace{1.0ex}
\noindent
\textbf{Classification Pipeline}. The next ablation study is to fix the sensor but change the entire classification pipeline. This will tell us how important the classifier is, and which classifier is more effective. The results in \fref{fig:results}(a) show that the among the competing methods, Dirty Pixels is the most promising one because it is end-to-end trained. However, comparing Dirty Pixels with our proposed method, at 1 ppp Dirty Pixels (shallow) achieves an accuracy of 53.9\% whereas the proposed (shallow) achieves 62.7\%. The trend continues as the photon level increases. This ablation analysis shows that even if we have a good sensor (QIS), we still need a good classifier.

\vspace{1.0ex}
\noindent
\textbf{Student-Teacher Learning}. Let us fix the sensor and the network, but change the training protocol. This will reveal the significance of the proposed student-teacher learning. To conduct this ablation study, we recognize that Dirty Pixels network structure (shallow and deep) is exactly the same as Ours (shallow and deep) since both use the same UNet and VGG-16. The only difference is the training protocol, where ours uses student-teacher learning and Dirty Pixels is a simple end-to-end. The result of this study is summarized in \fref{fig:results}(a). It is evident that our training protocol offers advantages over Dirty Pixels.

\begin{figure}[h]
\centering
\includegraphics[width = 1.0 \linewidth]{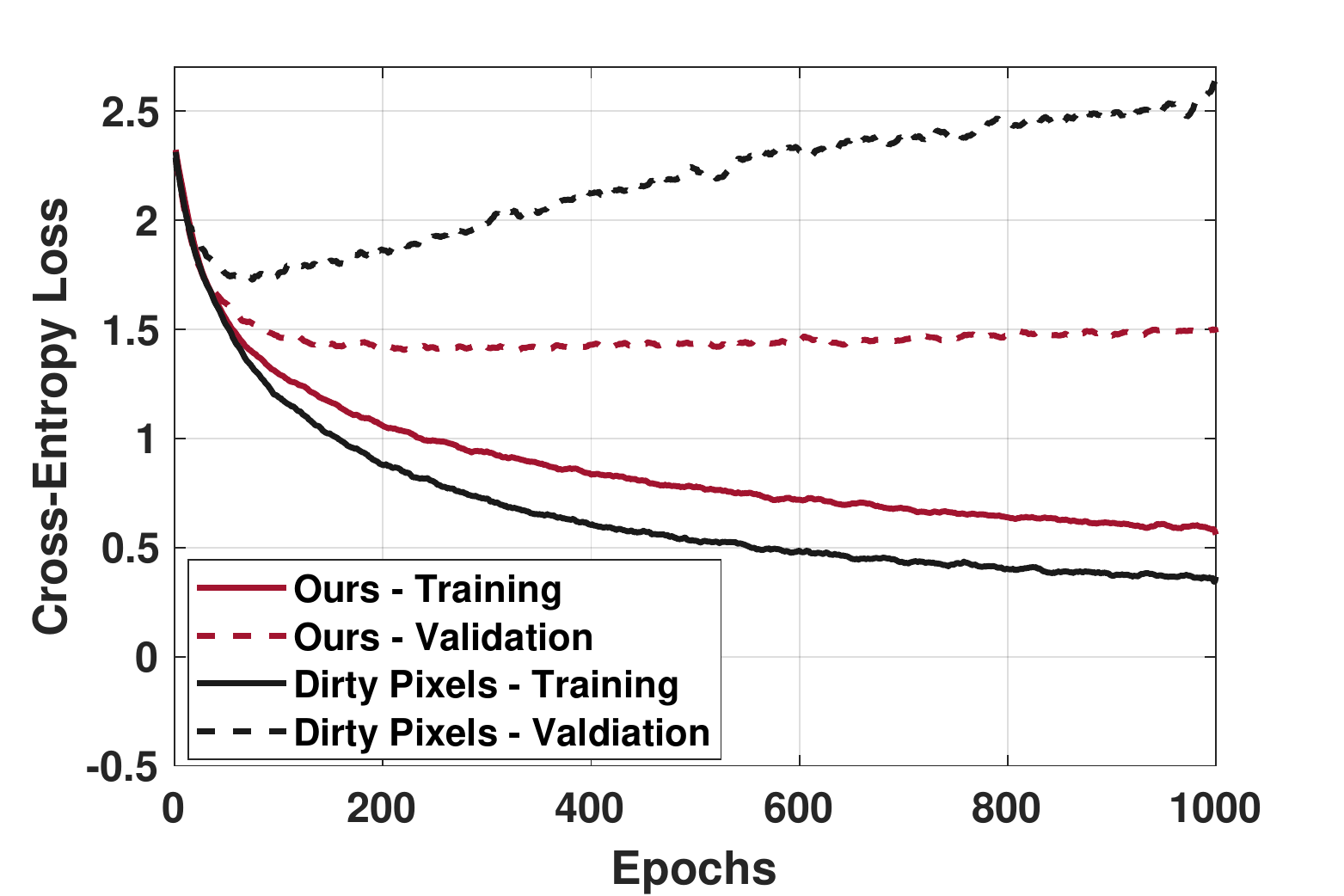}
\hspace{-2ex}
\caption{Training and validation loss of Our method and Dirty Pixels. Notice that while our training loss is higher, the validation loss is significantly lower than Dirty Pixels. }
\label{fig:valid_train_loss}
\end{figure}

We can further analyze the situation by plotting the training and validation error. \fref{fig:valid_train_loss} shows the comparison between the proposed method (shallow) and Dirty Pixels (shallow). It is evident from the plot that without stident-teacher learning (Dirty Pixels), the network overfits. This is evident if we look at the the validation loss, which drops and then rises whereas the training loss keeps dropping. In contrast, the proposed method appears to mitigate the overfitting issue. One possible reason is that the student-teacher learning is providing some kind of regularization in an implicit form so that the validation loss is maintained at a low level.

\vspace{1.0ex}
\noindent
\textbf{Choice of Classification Network}. All experiments reported in this paper use VGG-16 as the classifier. In this ablation study we replace the VGG-16 classifier by other popular classifiers, namely ResNet50 and InceptionV3. These networks are fine-tuned using the QIS data. Table~\ref{fig:valid_train_loss table} shows the comparisons. Using the baseline training scheme, i.e., simple fine-tuning as in Dirty Pixels, it is observed that there is a minor gap between the different classifiers. However, by using the proposed student-teacher training protocol, we observe a substantial improvement for all the classifiers. This ablation study confirms that student-teacher learning is not limited to a particularly network architecture.

\begin{table}[h]
\caption{Ablation study of different classifiers and different training schemes. Reported numbers are based on QIS synthetic experiments at 0.25 ppp for the Dog Dataset.}
  \begin{tabular}{cccc}
   \hline
   \hline
                & \hspace{1ex} VGG \hspace{1ex} & \hspace{1ex} ResNet \hspace{1ex} & Inception \\
    \hline
    Baseline    & 42.1\% & 43.3\%    & 44.3\% \\
    (fine-tuning) & &  & \\
    \hline
    Proposed    & 48.6\% & 49.1\%    & 50.0\% \\
    (student-teacher) & & & \\
    \hline
    Gap         &+6.5\%  & +5.8\%    & +5.7\% \\
    \hline
    \end{tabular}
\label{fig:valid_train_loss table}
\vspace{-2ex}
\end{table}

\vspace{1.0ex}
\noindent
\textbf{Using a pre-trained classifier}. This ablation study analyzes the effect of using a pre-trained classifier (trained on clean images). If we do this, then the overall system is exactly the same as the Restoration network \cite{liu2018connecting} in \fref{fig:methods_comparison}(c). Restoration network has three training losses: (i) MSE to measure the image quality, (ii) Perceptual loss to measure feature quality, and (iii) the cross-entropy loss. These three losses are used to just train the denoiser, and not the classifier. Since the classifier is fixed, it becomes necessary for the the denoiser to produce high-quality images or otherwise the classifier will not work. The results in \fref{fig:results}(a) suggest that when photon level is low, the denoiser fails to produce high quality images and so the classification fails. For example, at 0.25 ppp Restoration Network achieves 35.6\% but our proposed method achieves 52.1\%. Thus it is imperative that we re-train the classifier for low-light images.

\vspace{1.0ex}
\noindent
\textbf{Deep or Shallow Denoisers?} This ablation study analyzes the impact of using a deep denoiser compared to a shallow entrance layer. The result of this study can be found by comparing Ours (deep) and Ours (shallow) in \fref{fig:results}(a), as well as Dirty (deep) and Dirty (shallow). In both methods, the deep version refers to using a 20-layer UNet, whereas the shallow version refers to using a 2-layer network. The result in \fref{fig:results}(a) suggests that while the deep denoiser has significant impact to Dirty Pixels, its influence is quite small to the proposed method with the QIS images. One reason is that since we are using student-teacher learning, the features are already properly denoised. The added benefit from a deep denoiser for QIS is therefore marginal. However, for CIS data at low light, the deep denoiser actually helps getting better classification performance, especially when the signal level is much lower than the read noise.

\section{Conclusion}
We proposed a new low-light image classification method by integrating Quanta Image Sensors (QIS) and a novel student-teacher training protocol. Experimental results confirmed that such combination is effective for low-light image classification, and the student-teacher protocol is a better alternative than the traditional denoise-then-classify framework. This paper also made the first demonstration of low-light image classification at a photon level of 1 photon per pixel or lower. The student-teacher training protocol is transferable to conventional CIS data, however to achieve the desired performance at low light, it is necessary for QIS to be part of the overall pipeline. Using multiple frames for image classification would be a fruitful direction for future work.

\bibliographystyle{IEEEbib}
\bibliography{egbib}
\end{document}